\documentclass{ws-ijgmmp}
\usepackage{amssymb}
\usepackage{amsmath}

\begin{document}

\markboth{Fabio Briscese and Francesco Calogero} {Isochronous
Cosmologies}

\catchline{}{}{}{}{}
%
%%%%%%%%%%%%%%%%%%%%%%%%%%%%%%%%%%%%%%%%%%%%%%%%%%%%%%%%%%%%%%%%%%%%

\title{ISOCHRONOUS COSMOLOGIES}

\author{FABIO BRISCESE}

\address{Istituto Nazionale di Alta Matematica Francesco Severi, Gruppo
Nazionale di Fisica Matematica, Citt\`{a} Universitaria, P.le A.
Moro 5, 00185 Rome, Italy. \\
and\\
Dipartimento SBAI, Sezione di Matematica, Sapienza
Universit\`{a} di Roma, Via Antonio Scarpa 16, 00161 Rome, Italy.
\\
\email{fabio.briscese@sbai.uniroma1.it}}

\author{FRANCESCO CALOGERO}

\address{Dipartimento di Fisica, Universit\`{a} di Roma \textquotedblleft La
Sapienza\textquotedblright , Rome, Italy.\\
and\\
Istituto Nazionale di Fisica Nucleare, Sezione di Roma, Rome,
Italy.
\\
\email{francesco.calogero@roma1.infn.it,
francesco.calogero@uniroma1.it} }

\maketitle

\begin{history}
\received{(20 January 2014)} \revised{(04 February 2014)}
\end{history}

\begin{abstract}
The possibility has been recently demonstrated to manufacture
(nonrelativistic, Hamiltonian) many-body problems which feature an
isochronous time evolution with an arbitrarily assigned period $T$
yet mimic with good approximation, or even exactly, any given
many-body problem (within a large, physically relevant, class)
over times $\tilde{T}$ which may also be arbitrarily large (but of
course such that $\tilde{T}<T$). Purpose and scope of this paper
is to explore the possibility to extend this finding  to a general
relativity context. For simplicity we restrict our consideration
to the case of homogeneous and isotropic metrics and show that,
via an approach analogous to that used for the nonrelativistic
many-body problem, a class of homogeneous and isotropic
\textit{cyclic} solutions of Einstein's equations may be obtained.
For these solutions the duration of the cycles does not depend on
the initial conditions, so we call these models
\textit{isochronous cosmologies}. We give a physical
interpretation of such metrics and in particular we show that they
may behave \textit{arbitrarily} closely, or even
\textit{identically}, to the Friedman-Robertson-Walker solutions
of Einstein's equations for an \textit{arbitrarily} long time (of
course shorter than their period, which can also be assigned
\textit{arbitrarily}), so that they may reproduce all the
satisfactory phenomenological features of the standard
cosmological $\Lambda $-CDM model in a portion of their cycle;
while these isochronous cosmologies may be \textit{geodesically
complete} and therefore \textit{singularity-free}.
\end{abstract}

\keywords{Cosmology; Isochronous Systems; General Relativity.}

\section{Introduction}

It has been recently shown \cite{CL2007,calogero} how, given a quite general
(autonomous) dynamical system $D$, other (also autonomous) dynamical systems
$\tilde{D}$ can be manufactured, featuring two additional \textit{arbitrary}
positive parameters $T$ and $\tilde{T}$ with $T>\tilde{T}$ (and possibly
also two additional dynamical variables) and having the following two
properties: (i) For the same variables of the original dynamical system $D$
the new dynamical system $\tilde{D}$ yields, over the time interval $\tilde{T%
}$, hence for an \textit{arbitrarily long} time, a dynamical
evolution which mimics \textit{arbitrarily closely} that yielded
by the original system $D$; up to corrections of order
$t/\tilde{T}$, or possibly even \textit{identically}. (ii) The
system $\tilde{D}$ is \textit{isochronous}: all its solutions (for
arbitrary initial data) are completely periodic with the assigned period $T$%
. A particularly interesting example of this phenomenon is the
standard Hamiltonian system describing, in an ambient space of
\textit{arbitrary} dimensions $d$ (including of course $d=3$), an
\textit{arbitrary} number $N$ of point particles with
\textit{arbitrary} masses, interacting among themselves via
potentials depending \textit{arbitrarily} from the particle
coordinates. Moreover it has been shown \cite{CL2007,calogero}
that, if this (autonomous) Hamiltonian $H$ is
translation-invariant (i. e., it features no external potentials),
new (also autonomous) Hamiltonians $\tilde{H}$ characterizing
\textit{modified} many-body problems can be manufactured that
feature the \textit{same} dynamical variables as $H$ (i. e., there
is then no need to introduce two additional dynamical variables)
and yield a time evolution \textit{quite close}, or even
\textit{identical} to that yielded by the
original Hamiltonian $H$ over the \textit{arbitrarily assigned} time $\tilde{%
T}$, while being \textit{isochronous} with the \textit{arbitrarily assigned}
period $T$ (of course with $T>\tilde{T}$).

The Hamiltonian model $H$ described above clearly encompasses a lot of
physics; and since it is difficult, or perhaps impossible, to distinguish
experimentally dynamical systems that behave \textit{arbitrarily closely},
or even \textit{identically}, over an \textit{arbitrarily long} period of
time, this finding---which is a proven mathematical theorem, at least for
realistic Hamiltonian many-body problems predicting a nonsingular future, as
it is natural in physics---has various remarkable implications for all those
who believe that physics is a science based on experimental verifications of
its laws. In particular it raises \cite{CL2007,calogero} interesting
questions about the distinction between integrable and nonintegrable
evolutions, the definition of chaotic behavior, the applicability of
statistical mechanics and the validity of the second principle of
thermodynamics (say, for $N\approx 10^{24}$), and about cosmology (say, for $%
N\approx 10^{85}$; including, in this case, issues which have an
eschatological connotation reminiscent of "eternal return" conceptions which
are, however, not our cup of tea).

The present paper is motivated by the potential relevance of the findings
outlined above for general relativity. Can some analogous finding be
obtained concerning cosmology, but in the context of general relativity
rather than classical (nonrelativistic) mechanics?

\textit{Remark 1.1}. Actually the finding described above can also be
extended to a quantum mechanical context, up to the ambiguities inherent in
the transition from classical to quantum mechanics (particularly significant
for Hamiltonians with a nonconventional kinetic energy component, as
featured by the modified Hamiltonians mentioned above) \cite{CL2007}. Anyway
in this paper we refrain from considering any quantum-mechanical context,
which would anyway be impossible as long as no framework is available
encompassing in a mathematically rigorous fashion both general relativity
and quantum theory. $\square $

Although the results outlined above are valid for autonomous systems, they
are in fact based on the introduction of an auxiliary variable in place of
the physical time variable \cite{calogero}. In particular the dynamical
systems $D$ and $\tilde{D}$ mentioned above feature the same trajectories in
phase space, but while the time evolution of the dynamical system $D$
corresponds to a \textit{uniform} forward motion along those trajectories,
the time evolutions of the modified dynamical systems $\tilde{D}$---although
produced by time-independent equations of motion---correspond to a \textit{%
periodic} (with assigned period $T$), forward and backward, time evolution
along those same trajectories, exploring of course only a portion of them;
with the possibility that, for a subinterval $\tilde{T}<T$, this motion be
also \textit{uniform}, entailing that on such subinterval the dynamics of $D$
and $\tilde{D}$ are indeed \textit{identical}. Let us emphasize that, while
this procedure entails the introduction of an auxiliary variable $\tau
\equiv \tau \left( t\right) $, appropriately related to the physical time $t$%
, any attempt to attribute to this new variable $\tau $ the significance of
"time" would be completely unjustified: quite improper and confusing.

In classical mechanics, these different behaviors are produced by \textit{%
different} (autonomous) Hamiltonians (some of which, as indicated
above, may however be hardly distinguishable by experiments
involving an arbitrarily long, but finite, time evolution). We
show below that, in the context of general relativity, an
analogous phenomenology also exists, which however does
\textit{not} require any change of the fundamental Einstein
equations, corresponding instead to the identification of an
enlarged class of metrics associated to the solutions of these
equations. This shall again entail the
introduction of a new auxiliary variable $\tau \equiv \tau \left( t\right) $%
---where $t$ is the physical time, itself defined up to diffeomorphic
transformations as implied by the general relativity context; and again any
attempt to attribute globally to this auxiliary variable $\tau $ the
significance of time would be completely unjustified, improper and confusing.

One last caveat before embarking in the detailed presentation of our
results. The previous findings concerning, in a nonrelativistic context, the
general many-body problem \cite{CL2007,calogero} were not meant to suggest
that the modified Hamiltonians yielding isochronous motions are more
appropriate than the standard Hamiltonians as descriptions of any specific
many-body problem---including that involving a number of particles
comparable to those of the entire universe (assuming this number has any
sense). It was merely meant to demonstrate a, perhaps unpleasant but
unfortunately inescapable, feature of any attempt to describe a physical
many-body problem in the context of Hamiltonian dynamics: the fact that
\textit{different} Hamiltonians exist which characterize \textit{different}
physical behaviors which may however be difficult, in fact impossible, to
distinguish experimentally. Of course this fact could be simply ignored, for
one's peace of mind; albeit at the risk of eventually discovering that the
pursuit of the peace of one's mind is not the proper approach to the
advancement of science (indeed, how to \textit{exclude} \textit{altogether}
the---presumably very unlikely, but not physically forbidden---possibility
that one of the modified Hamiltonians turn out to describe reality better
than the standard one?). Likewise the present paper---which is focussed on
cosmology and in that context shows that the framework of general relativity
includes a larger class of cosmological solutions than usually considered
and that these solutions may describe \textit{isochronous} universes---is
not meant to argue that our Universe \textit{does} evolve \textit{%
isochronously}; but merely that such possibilities are contained in the
equations of general relativity---compatibly with what is observationally
known about our Universe---and might indeed have some interesting
connotations, for instance avoid Big Bang singularities.

For simplicity in this paper we focus on cosmological solutions of the
Einstein equations providing a highly idealized (\textit{homogeneous} and
\textit{isotropic}) picture of the universe; which is indeed the standard
point of departure of cosmological investigations (see for instance \cite%
{mukhanov}). Applying techniques analogous to those outlined above \cite%
{CL2007,calogero}, we obtain \textit{isochronous} (\textit{homogeneous} and
\textit{isotropic}) cosmological solutions of Einstein's equations and
tersely analyze their properties.

Periodic cosmological solutions of Einstein's equations have already been
abundantly discussed as alternatives to the standard cosmological $\Lambda $%
-CDM model consisting (see for instance \cite{mukhanov}) of a
Friedmann-Robertson-Walker (FRW) homogeneous and isotropic universe filled
with standard particles plus a cosmological constant $\Lambda $ plus some
amount of dark matter. Let us recall in this connection that---in spite of
the fact that the $\Lambda $-CDM model is rather successful at explaining
the data \cite{cmb,leansing,bao,lss,supernovae,ref0,ref1,ref2,ref3,ref4}%
---some questions remain open.  For example the the appropriate
value of the cosmological constant cannot be deduced from a
fundamental theory, due to a severe fine tuning problem.  And it
is unclear which inflaton field should be associated to the primordial inflation. Moreover the $\Lambda $%
-CDM model features a Big Bang singularity and requires a fine tuning
between the energy density of matter, radiation and dark energy as well as
specific initial conditions.

An example of cyclic universe makes use of a scalar field with a specific
potential such that the universe starts from a big bang and ends with a big
crunch \cite{cyclic}. Other examples are the ekpyrotic scenario \cite{branes}
based on brane theory, the string theory inspired bouncing cosmologies \cite%
{bouncing} and the periodic cosmologies based on Chaplygin gas \cite%
{chaplyging}. All these models make use of some string theory inspired
scalar or vector field, or of a perfect fluid with a nonlinear equation of
state (see \cite{bouncingrewiev} for a review), resulting in a FRW universe
filled with some exotic fluid.

In the present paper we extend to general relativity the results on \textit{%
isochronous} systems valid in the context of nonrelativistic Hamiltonian
many-body problems. We thus obtain a class of \textit{cyclic} solutions of
Einstein's equations which are \textit{homogeneous} and \textit{isotropic}
in space. These solutions are cyclic even if they contain no exotic fluids
but only ideal fluids. Since the duration of each cycle does not depend on
initial conditions we call them \textit{isochronous} cosmological solutions.
Another remarkable property of these solutions is that they can mimic the
FRW solutions of Einstein's equations accurately for an arbitrarily long
time. Therefore they are no less in agreement with cosmological data than
the $\Lambda $-CDM model, giving the right sequence of inflation / radiation
domination / matter domination / late time acceleration epochs. Moreover,
they may be free of singularities: the big bang / big crunch singularities
may be avoided because in these solutions the contraction and expansion of
the universe may be reversed at some finite (neither vanishing nor
diverging) value of the scale factor. Therefore there is no need to explain
the transition from a big crunch to a big bang as is the case for other
cyclic universes \cite{cyclic}.

Let us conclude this preliminary section by noting that the solutions of
Einstein's equations introduced in this paper entail metrics that are
considered unacceptable by some colleagues: indeed we experienced
difficulties in having this paper published, because some referees ruled out
these solutions as \textit{unphysical}. This opinion was generally based on
the unjustified interpretation of the quantity $\tau \equiv \tau \left(
t\right) $ (see below) as a "global time", rather than an auxiliary function
of the time $t$ (as in the nonrelativistic case described above and treated
in \cite{calogero}). In fact, as shown below, the only peculiarity of the
metrics we consider is that they are \textit{degenerate} at some discrete
values of the parameter $t$ which---up to diffeomorphic
reparameterization---has in our models the significance of "time"; a feature
that is also shared by other standard solutions of Einstein's equations
which are nevertheless generally considered quite acceptable, such as, for
instance, the well-known Schwarzschild solution. The arguments of these
colleagues seemed to us as unyielding to our efforts to clarify matters as
Cesare Cremonini's refusal to look into Galileo's telescope \cite{Cremonini}%
. We hope that the publication of this paper will allow the scientific
community to judge whether this extension of the class of solutions of
Einstein's equations should be considered acceptable and the corresponding
physical arguments worthy of serious consideration (incidentally, we will be
happy to share with any interested reader our previous exchanges with
referees and editors). Let us reiterate that we are not asserting in this
paper that the class of solutions we introduce provide \textit{the} correct
(of course approximate) description of the Universe we live in; but we
submit that they should be taken into proper consideration, as they seem to
us to feature---as indicated above and below---certain, possibly more
appealing, aspects than alternative descriptions in the framework of general
relativity of an (\textit{isotropic} and \textit{homogeneous}) \textit{%
isochronous} Universe. In any case it seems to us that one should not ignore
the possibility to construct---as shown below--- for any given solution of
the Einstein equations, another solution---in fact, an infinity of such
solutions---which are physically essentially indistinguishable from the
given one over some (arbitrarily long) time interval but are periodic on a
longer time scale.

\bigskip

\section{Isochronous Cosmologies}

Our point of departure is to consider, in a given reference frame $(t,\vec{x}%
)$, an homogeneous and isotropic metric reading
\begin{subequations}
\label{NEWmetric}
\begin{equation}
ds^{2}=b(t)^{2}~dt^{2}-a(t)^{2}~d\vec{x}^{2}  \label{newmetric}
\end{equation}%
with $b\left( t\right) $ a \textit{periodic} function (with an \textit{%
arbitrarily assigned} period $T$) having moreover a vanishing mean value, so
that its integral $B\left( t\right) $ is also periodic with period $T$:%
\begin{equation}
b(t+T)=b(t)~;~~~~B\left( t\right) \equiv \int_{0}^{t}b\left( t^{\prime
}\right) dt^{\prime }~,~~~B\left( t+T\right) =B\left( t\right) ~.
\label{bBperiodic}
\end{equation}

Of course this metric can be mapped \textit{locally} into the FRW metric
\end{subequations}
\begin{subequations}
\label{FRWmetric}
\begin{equation}
ds^{2}=d\tau ^{2}-\alpha (\tau )^{2}~d\vec{x}^{2}~,~~~\alpha (\tau )\equiv
a(t)~,  \label{FRWmetric1}
\end{equation}%
via the change of variable
\begin{equation}
d\tau =b(t)~dt~,~\hspace{0in}~\tau \left( t\right) =B\left( t\right) ~.
\label{repar}
\end{equation}%
Hence one might infer that the new metric (\ref{NEWmetric}) is physically
equivalent to the FRW metric (\ref{FRWmetric}), since in general relativity
different metrics related by a reparameterization of the coordinates---and
in particular of the time variable---are considered physically equivalent.
But this is the case only if the relevant reparameterization is \textit{%
univocally invertible} (technically, a \textit{diffeomorphism}). This is
indeed the case for the change of variables (\ref{repar}) \textit{locally},
but \textit{not globally}; in particular, certainly \textit{not over a time
interval equal to or larger than }$T$, because obviously the periodic
functions $b(t)$ and $B\left( t\right) $ are certainly \textit{not
univocally invertible over time intervals equal to or larger than }$T$ (see (%
\ref{bBperiodic}); in fact, here and below $T$ should be replaced by $T/2$
for the assignment (\ref{blambda}), see below). Hence the two metrics (\ref%
{NEWmetric}) and (\ref{FRWmetric}) are \textit{not physically equivalent}:
they represent two \textit{physically different} solutions of the Einstein
equations. Let us reemphasize this point: the metric (\ref{NEWmetric}) is
\textit{not} globally reducible via a diffeomorphism to the FRW metric (\ref%
{FRWmetric}) hence the class of solutions of the Einstein's equations
characterized by this metric (\ref{NEWmetric}) represent a physically
different universe from those characterized by the FRW metric (\ref%
{FRWmetric}). In a given spacetime described by the metric (\ref{NEWmetric})
an observer is allowed to change the way time is measured---i. e., to
perform a diffeomorphic transformation of the time variable---in order to
reduce the spacetime metric to the FRW metric (\ref{FRWmetric}); but only
for an interval of time $\tilde{T}$ sufficiently smaller than $T$. Over any time interval larger than $T$%
---or even insufficiently smaller than $T$, see below---a departure from the
FRW metric (\ref{FRWmetric}) shall necessarily emerge. Indeed we show below
that the new solutions associated to the metric (\ref{NEWmetric}) are
\textit{physically different} from those associated to the metric (\ref%
{FRWmetric}): for instance in the first case---in contrast to the
second---there might be no Big Bang (see below). And we also indicate why
there is no justification to exclude \textit{a priori} these solutions.

As usual in cosmological investigations based on the Einstein equations, we
assume that the energy-momentum tensor of the universe is that of an ideal
fluid with a 4-velocity given by $U_{\mu }=|b(t)|~\delta _{\mu 0}$ (the
extension to more general energy-momentum tensors is trivial), so that the
ideal fluid has zero spatial velocity and the frame $(t,\vec{x})$ is
co-moving, while the zero component of the 4-velocity is always nonnegative
(as it should be). The Einstein equations for the metric (\ref{NEWmetric})
are then drastically simplified \cite{mukhanov}. The corresponding
generalized FRW equations read
\end{subequations}
\begin{subequations}
\label{3}
\begin{equation}
3\left[ \frac{\dot{a}(t)}{a(t)}\right] ^{2}=k\,b(t)^{2}\rho (t)~,
\label{NEWFRW1}
\end{equation}%
\begin{equation}
\dot{\rho}(t)+3\left[ \frac{\dot{a}(t)}{a(t)}\right] \left[ \rho (t)+P(t)%
\right] =0~.  \label{NEWFRW3}
\end{equation}%
Here and hereafter superimposed dots denot differentiation with respect to
the time $t$; $\rho $ and $P$ are respectively the energy density and
pressure of the universe; $k=8\pi G/c^{4}$; $G$ is the gravitational
constant; $c$ the speed of light. We stress that these equations are
obtained in the framework of Einstein's theory and they do not assume any
modification of general relativity.

This system of two ODEs must be complemented by an equation of state
relating the pressure $P$ of the universe to its energy density $\rho $: we
shall here use for simplicity the simple ideal fluid relation $P=\omega \rho
$, with constant $\omega $. But note that, after this assignment, the system
(\ref{3}) is not quite closed: it is still possible to assign, essentially
arbitrarily, the function $b(t)$. Of course this arbitrariness corresponds
to the freedom in general relativity to reparametrize time; but---as
discussed above---it goes beyond this if $b\left( t\right) $ is \textit{not}
invertible, hence if the relevant reparameterization is \textit{not} a
global diffeomorphism.

If we consider a perfect fluid with constant equation of state parameter $%
\omega >-1$, it follows from (\ref{NEWFRW3}) that the energy density of the
fluid is
\end{subequations}
\begin{equation}
\rho (t)=\rho \left( 0\right) \left[ \frac{a(0)}{a(t)}\right] ^{3(\omega
+1)}~,~~~\text{if~~~}\omega >-1~,  \label{FRWrho}
\end{equation}%
while for $\omega =-1$ one has a constant energy density,%
\begin{equation}
\rho =\Lambda ~~~\text{if \ \ }\omega =-1~.
\end{equation}%
It is now convenient \cite{calogero} to introduce a new variable $%
\tau $ by setting (see (\ref{bBperiodic}))
\begin{subequations}
\label{6}
\begin{equation}
\tau (t)\equiv B\left( t\right) ~,  \label{tau}
\end{equation}%
and new auxiliary functions $\alpha \left( \tau \right) ,~r\left( \tau
\right) ,~p\left( \tau \right) $ by setting
\begin{equation}
a(t)\equiv \alpha \left( \tau (t)\right) ~,~~~\rho (t)\equiv r\left( \tau
(t)\right) ~,~~~P(t)\equiv p\left( \tau (t)\right) ~.  \label{definitions}
\end{equation}%
Then the functions $a(t)$ and $\rho (t)$ are solutions of (\ref{3}) provided
$\alpha (\tau )$ and $r(\tau )$ are solutions of the following system:
\end{subequations}
\begin{subequations}
\label{10}
\begin{equation}
3\left[ \frac{\alpha ^{\prime }(\tau )}{\alpha (\tau )}\right] ^{2}=kr(\tau
)~,  \label{FRW1}
\end{equation}%
\begin{equation}
r^{\prime }(\tau )+3\left[ \frac{\alpha ^{\prime }(\tau )}{\alpha (\tau )}%
\right] \left[ r(\tau )+p(\tau )\right] =0~.  \label{FRW3}
\end{equation}%
Here the prime denotes differentiation with respect to $\tau $, and $p(\tau
) $ and $r(\tau )$ satisfy the same equation of state $p(\tau )=\omega
\,r(\tau )$ as $P(t)$ and $\rho (t)$. And since the definition of $\tau (t)$
(see (\ref{tau}) with (\ref{bBperiodic})) implies that this function is
periodic with period $T$,
\end{subequations}
\begin{subequations}
\begin{equation}
\tau (t+T)=\tau (t)~,
\end{equation}%
the same property is inherited via the definitions (\ref{definitions}) by
the physical quantities $a(t)$, $\rho (t)$ and $P(t)$:%
\begin{equation}
a(t+T)=a(t)~,~~~\rho (t+T)=\rho (t)~,~~~P(t+T)=P(t)~.
\end{equation}%
Note that this property holds now for the solutions of (\ref{3})
characterized by any initial conditions: the physical quantities $a(t)$, $%
\rho (t)$ and $P(t)$ are therefore \textit{isochronous}. Moreover, since $%
b(t)$ can be assigned arbitrarily, the period $T$ of these \textit{%
isochronous} solutions is a parameter which can be freely assigned.

But let us re-emphasize that the variable $\tau $ cannot be given \textit{%
globally} the significance of "time", hence the fact that $\dot{\tau}$ might
change sign---indeed, it certainly does so, see (\ref{tau}) and (\ref%
{bBperiodic})---has no \textit{unphysical }connotation.

It is immediate to identify (\ref{10}) with the FRW system characterized by
the metric (\ref{FRWmetric1}) with the energy momentum tensor of a perfect
fluid in co-moving coordinates. Therefore (\ref{6}) says that the solutions $%
a(\tau )$, $\rho (\tau )$, $p(\tau )$ of the FRW system (\ref{10}) are
\textit{locally} mapped into the solutions of (\ref{3}) via the \textit{local%
} time reparameterization (\ref{6}). But, as emphasized above, due to the
periodicity of $\tau \left( t\right) $, this mapping cannot be globally
extended since the change of variable (\ref{6}) is not a diffeomorphism for
all $t$, therefore $a(t),\rho (t),P(t)$ and $\alpha (\tau ),r(\tau ),p(\tau
) $ correspond in fact to \textit{different} cosmologies.

Below we present two explicit examples of isochronous cosmological solutions
of Einstein's equations obtained in this manner and describe their main
properties. In both cases we assign for simplicity the function $b(t)$ and $%
B\left( t\right) $ as follows:
\end{subequations}
\begin{equation}
b(t)\equiv \cos \left( \Omega t\right) ~,~~~B\left( t\right) =\frac{\sin
\left( \Omega t\right) }{\Omega }~,~~~\Omega =\frac{2\pi }{T}\ .
\label{blambda}
\end{equation}

As first example, consider a universe filled with a dark energy fluid with
constant $\rho =\Lambda $ and equation of state parameter $\omega =-1$.
Assuming that the metric tensor is given by (\ref{NEWmetric}), the system (%
\ref{3}) has the following solution:%
\begin{equation}
a(t)=a_{0}\exp \left[ \sqrt{\frac{k\Lambda }{3}}~\frac{\sin \left( \Omega
t\right) }{\Omega }\right] ~,~~~a_{0}\equiv a\left( 0\right) ~.
\label{alambda}
\end{equation}%
Note that at the times $t_{n}=(1/2+n)\pi /\Omega $ with $n$ integer the
scale factor $a(t)$ reaches its maximum (for $n$ even) and minimum (for $n$
odd) values $a_{\pm }\equiv a_{0}\exp \left[ \pm \sqrt{k\Lambda /3\Omega ^{2}%
}\right] $. Also note that, even if the metric (\ref{NEWmetric}) is
degenerate since $b(t_{n})=0$, all physical quantities as the energy density
and pressure, the Ricci scalar curvature $R$, etc., are \textit{not}
singular since they can be expressed as functions of the scale factor $%
a\left( t\right) $ (e. g. $R=k\left[ \rho (a)-3P(a)\right] $), which is
finite for all time, see (\ref{alambda}). Therefore $t_{n}$ is a fictitious
singularity corresponding to the time when the universe passes from an
expanding to a contracting epoch and viceversa.

As second example, consider the case of a perfect fluid with equation of
state parameter $\omega >-1$. Then the scale factor is

\begin{equation}
a(t)=a_{0}~\left[ 1+(1+\omega )\sqrt{\frac{3k\rho _{0}}{4}}\frac{\sin \left(
\Omega t\right) }{\Omega }\right] ^{\frac{2}{3(1+\omega )}}~,~~~a_{0}\equiv
a\left( 0\right) ~.  \label{aomega}
\end{equation}

The maximum and minimum of the scale factor are now again reached at $%
t_{n}=(1/2+n)\pi /\Omega $ and have the values $a_{\pm }=a_{0}\left[ 1\pm
\sqrt{3(1+\omega)^{2}k\rho _{0}/4\Omega ^{2}}\right] ^{\frac{2}{%
3(1+\omega )}}$. Also in this case $t_{n}$ corresponds to the transition
from the expanding to the contracting phases and viceversa and to a
fictitious singularity of the metric, since $b(t_{n})=0$ but the scale
factor $a(t)$ and the relevant physical quantities may remain finite for all
time. Indeed for any assignment of the parameter $\Omega $ such that%
\begin{equation}
\Omega >\sqrt{\frac{3k\rho _{0}}{4}}~\left( 1+\omega \right) ~,
\label{nobigbang}
\end{equation}%
the scale factor $a(t)$ is positive for all time (see (\ref{aomega})) and
the big bang singularity of the FRW model is avoided. Hereafter we assume
that this restriction, (\ref{nobigbang}), is always enforced.

Of course with a different assignment of $b(t)$ than (\ref{blambda}) one
obtains different solutions of (\ref{3}). The simple assignment (\ref%
{blambda}) is an interesting example inasmuch as the scale factor $a(t)$
solution of (\ref{3}) then mimic the scale factor $\alpha (\tau )$ of the $%
\Lambda $-CDM model over time intervals much shorter than $T$, see (\ref%
{blambda}). Indeed in this case, as long as $|t|\ll T$ one has
$b(t)\simeq 1$ and $\tau \simeq t$ and therefore $a(t)\simeq
\alpha (t)$. Since $\Omega $ hence $T$ is arbitrary (see
(\ref{blambda})) one can for instance choose $T$ just a bit
smaller than the "standard age" of the universe, say $T \lesssim
1/H_{0}$ where $H_{0}\simeq 70Km/sMpc$ is the current value of the
Hubble parameter.

We stress that the argument based on the transition in the reduced Einstein
equations from the time $t$ to the auxiliary variable $\tau $, hence from (%
\ref{3}) to (\ref{10})---illustrated above when the right-hand side is
characterized by the simple ideal fluid relation $P=\omega \rho $---remains
valid in the more general case in which the right-hand side of (\ref{3})
corresponds to a universe filled with a mixture of ultra-relativistic and
nonrelativistic perfect fluids plus a cosmological constant and an inflaton
scalar field; then the solution of (\ref{3}) and (\ref{10}) gives, for $%
|t|\ll T$, the "right" sequence of inflation / radiation domination / matter
domination / late time acceleration epochs characterizing the $\Lambda $-CDM
universe. While, as discussed above, a suitable assignment of $\Omega $
hence $T$ allows to evade the big bang singularity (see (\ref{nobigbang}) in
the case of a perfect fluid with $\omega >-1$).

\bigskip

\section{Generalities on isochronous cosmologies}

We complete this paper with some general considerations on these \textit{%
isochronous} cosmological models. We have shown how to construct isochronous
solutions of (\ref{3}) from solutions of the FRW equations (\ref{10}) by
considering arbitrary periodic functions $b(t)$ such that their integral $%
B\left( t\right) $ is also periodic and the big bang singularity is avoided.
In such a way one obtains solutions which are also periodic with the same
period of $b(t)$ in the frame $(t,\vec{x})$. Their period is independent of
the initial data hence these solutions are \textit{isochronous}. We call
these periodic and singularity free models \textit{isochronous cosmologies},
to distinguish them from other cyclic models \cite{bouncingrewiev}.

Let us reemphasize that within any time interval of length appropriately
less than $T$ (in particular, less than $T/2$ for the specific assignment (%
\ref{blambda})) where the change of the time variable (\ref{6}) is
invertible, the isochronous solutions based on the metric (\ref{NEWmetric})
can be mapped into FRW solutions based on the metric (\ref{FRWmetric}) by
the time reparameterization $t\rightarrow \tau (t)=B\left( t\right) ,$ see (%
\ref{tau}). This implies that \textit{isochronous} solutions are \textit{%
locally} (in time) equivalent to FRW solutions in the sense that they give
the same physics. However, since the change of time (\ref{6}) is \textit{not}
a global diffeomorphism, the equivalent FRW frame is \textit{not} a global
system of coordinates, and the equivalence is only \textit{local} (in time).
In fact the isochronous solutions are such that the scale factor $a(t)$ is
not a monotonous function of time; moreover the big bang singularity is
avoided, a fundamental difference from the FRW cosmology. Isochronous and
FRW solutions are locally but not globally equivalent.

From (\ref{6}) it is easy to recognize that the isochronous solutions $%
a(t),~\rho (t),~p(t)$ span only a part of the FRW trajectories $\alpha (\tau
),~r(\tau ),~P(\tau )$ and the corresponding \textit{isochronous} universes
consist of an infinite sequence of expansions and contractions. The turning
points between each expanding and contracting phase (and vice versa) occur
at the times $t_{n}$ when $b(t)$ changes sign and the metric (\ref{NEWmetric}%
) is degenerate since $b(t_{n})=0$.

Let us however note that, while the metric (\ref{NEWmetric}) is indeed
degenerate on the hypersurfaces $t=t_{n}$ since $g_{00}=b(t_{n})^{2}=0$ and $%
g_{\mu \nu }$ is not invertible, these hypersurfaces do \textit{not}
correspond to physical singularities, since $a(t_{n})$ neither vanishes nor
diverges; hence the energy density $\rho (t)$ and pressure $p(t)$ of the
universe are finite at $t=t_{n}$ as well as the Ricci scalar curvature $R=-k%
\left[ \rho (t)-3P(t)\right] $. In fact, by use of the first of (\ref%
{definitions}) one has
\begin{equation}
R=\frac{a\ddot{a}b-a\dot{a}\dot{b}+\dot{a}^{2}b}{a^{2}b^{3}}=\frac{\alpha
(\tau (t))\alpha ^{\prime \prime }(\tau (t))+\alpha ^{\prime 2}}{\alpha
(\tau (t))^{2}}~,
\end{equation}%
where, above and below, superimposed dots denote as usual
differentiations with respect to the time $t$\ and appended primes
denote differentiations with respect to the argument of the
function they are appended to, so that $\alpha ^{\prime
}(z)=d\alpha (z)/dz$ and $\alpha ^{\prime \prime }(z)=d^{2}\alpha
(z)/dz^{2}$. One can also compute the Kretschmann invariant
$K\equiv R^{\alpha \beta \gamma \theta }R_{\alpha \beta \gamma
\theta }$ which turns out to be expressed by the following
formulas:
\begin{equation}
K=\frac{12\left[ a^{2}\ddot{a}^{2}b^{2}-2a^{2}\dot{a}\ddot{a}b\dot{b}+a^{2}%
\dot{a}^{2}\dot{b}^{2}+\dot{a}^{4}b^{2}\right] }{a^{4}b^{6}}=\frac{12\left[
\alpha (\tau (t))^{2}\alpha ^{\prime \prime }(\tau (t))+\alpha ^{\prime 4}%
\right] }{\alpha (\tau (t))^{4}}~.
\end{equation}%
Therefore both the Ricci scalar and the Kretschmann invariant remain \textit{%
finite} for all time $t$. Let us also mention that $R_{ii}\propto
\rho (t)$, $R_{00}\propto b(t)^{2}\rho (t)$ and $T_{00}\propto
b(t)^{2}\rho (t)$ are also \textit{finite} at $t_{n}$, while
$R^{00}\propto \rho (t)/b(t)^{2}$ and $T^{00}\propto \rho
(t)/b(t)^{2}$ are \textit{not}. Physical singularities are
characterized by the fact that scalar quantities such as the
energy density, the pressure and the Ricci scalar curvature become
infinite on the singularity. The fact that these physical scalar
quantities remain instead \textit{finite} at $t=t_{n}$ indicates
that the hypersurfaces $t=t_{n}$ are \textit{not} physical
singularities.

Let us further elaborate this point and show that the spacetime with metric (%
\ref{NEWmetric}) is not singular at $t=t_{n}$ and more generally that, with
a proper choice of the function $b(t)$, the isochronous solutions are
singularity-free, see for instance the case of a perfect fluid with equation
of state parameter $\omega $ with the assignment (\ref{blambda}) and the
condition (\ref{nobigbang}).

Indeed a spacetime is \textit{singular} if it is \textit{not geodesically
complete}; a spacetime $M$ is geodesically complete if, for every point $q$
of $M$, the exponential map $\exp (q)$ is defined on the entire tangent
space $T_{q}M$ \cite{Nat} and geodesics are future- and past-extendible.

Let us suppose that the geodesics associated with the FRW metric (\ref%
{FRWmetric}) are identified by the formula%
\begin{equation}
Y =\left[ \lambda ,~\vec{Y}\left( \lambda \right) \right] ~,
\label{GeodesicFRW}
\end{equation}%
where the parameter $\lambda $ coincides with the FRW cosmological time $%
y^{0}=\tau $ along the geodesics. Since the FRW solution has a physical big
bang singularity at finite time, say at $\tau =\tau _{s}$, the geodesics (%
\ref{GeodesicFRW}) are defined for any $\lambda \geq \tau _{s}$ and are
past-inextendible to $\lambda <\tau _{s}$, in accordance with the fact that
the FRW spacetime is not geodesically complete.

It is easy to show (see Appendix A) that the geodesics associated with the
isochronous metric (\ref{NEWmetric}) are identified by the analogous formula%
\begin{equation}
X = \left[ \mu ,~\vec{Y}\left( B\left( \mu \right) \right) \right] =\left[
\mu ,~\vec{Y}\left( \tau \left( \mu \right) \right) \right]
\label{Geodesic isochronous 0}
\end{equation}%
and in this case the parameter $\mu $ coincides with the time $x^{0}=t$
along the geodesics of the isochronous spacetime. Therefore for any function
$b(t)$ such that
\begin{equation}
\tau (t)=B(t)>\tau _{s},\qquad \forall t\in \mathbb{R},  \label{condition b}
\end{equation}%
the geodesics associated with the isochronous metric (\ref{NEWmetric}) are
defined for any $t$ and are past- and future-extendible. These geodesics are
open spiraling curves in spacetime, with the space coordinates evolving
periodically as functions of the time coordinate $x^{0}=\mu $. It may
therefore be concluded that any spacetime associated with the isochronous
metric (\ref{NEWmetric}) with $b(t)$ fulfilling the condition (\ref%
{condition b}) is geodesically complete and therefore singularity free. This
in turn implies that the hypersurfaces $t=t_{n}$ are not physical
singularities of the isochronous solutions.

To give an explicit example, we return to the case of a spacetime filled
with a perfect fluid with equation of state parameter $\omega >-1/3$. The
FRW scale factor is then

\begin{equation}
\alpha (\tau )=a_{0}~\left[ 1+(1+\omega )\sqrt{\frac{3k\rho _{0}}{4}}\tau %
\right] ^{\frac{2}{3(1+\omega )}}~,~~~\alpha _{0}\equiv \alpha \left(
0\right)  \label{FRW a}
\end{equation}%
for any $\tau \geq \tau _{s}\equiv -\sqrt{4/3k\rho _{0}}/\left( 1+\omega
\right) $ with the big bang singularity at $\tau _{s}$. Let us consider for
instance the FRW light-like geodesics moving along the $Y^{1}$ direction,
which are given by
\begin{subequations}
\label{GeodesicFRW 1}
\begin{equation}
Y=\left[ \lambda ,Y^{1}\left( \lambda \right) ,Y_{s}^{2},Y_{s}^{3}\right]
\end{equation}%
with constant $Y_{s}^{2}$ and $Y_{s}^{3}$ and

\begin{equation}
Y^{1}(\lambda )=Y_{s}^{1}\pm \int_{\tau _{s}}^{\lambda }\frac{d\tau ^{\prime
}}{\alpha (\tau ^{\prime })}=Y_{s}^{1}\pm \frac{2}{\alpha _{0}(1+3\omega )}%
\sqrt{\frac{3}{k\rho _{0}}}\left[ 1+\frac{3(1+\omega )}{2}\sqrt{\frac{k\rho
_{0}}{3}}\lambda \right] ^{1-\frac{2}{3(1+\omega )}}~,  \label{FRW geod}
\end{equation}%
and are well defined for any $\lambda \geq \tau _{s}$. The corresponding
light-like geodesics of the associate isochronous metric will be given by (%
\ref{Geodesic isochronous 0}) as
\end{subequations}
\begin{subequations}
\label{Geodesic isochronous 1}
\begin{equation}
X=\left[ \mu ,X^{1}\left( \mu \right) ,Y_{s}^{2},Y_{s}^{3}\right]
\end{equation}%
with
\begin{equation}
X^{1}(\mu )=Y^{1}(B(\mu ))=Y_{s}^{1}\pm \frac{2}{\alpha _{0}(1+3\omega )}%
\sqrt{\frac{3}{k\rho _{0}}}\left[ 1+\frac{3(1+\omega )}{2}\sqrt{\frac{k\rho
_{0}}{3}}B(\mu )\right] ^{1-\frac{2}{3(1+\omega )}}  \label{FRW geod}
\end{equation}%
and $B(t)$ given by (\ref{6}). Therefore by assigning $b(t)$ so that (\ref%
{condition b}) is satisfied, e.g. with the assignment (\ref{blambda}), the
condition (\ref{condition b}) reads $\sin (\Omega t)/\Omega >-\sqrt{4/3k\rho
_{0}}/\left( 1+\omega \right) $ yielding the condition (\ref{nobigbang}).
Then the geodesics (\ref{Geodesic isochronous 1}) are defined for any $t$,
in accordance with the fact that with such a choice of $b(t)$, isochronous
metrics are geodesically complete and singularity free, see also (\ref%
{aomega}).

For completeness we also show that the isochronous metric (\ref{NEWmetric})
fulfills the junction conditions on the hypersurfaces $t=t_{n}$ and we
obtain the expression of the extrinsic curvature on the hypersurfaces $%
t=const$. We do so because one might question whether the
isochronous metric we have introduced is physically acceptable
and, in particular, whether the junction conditions on any
hypersurfaces---especially on the hypersurfaces $t=t_{n}$ where
the expansion / contraction phases alternate and the isochronous
metric is degenerate---are satisfied and the stress-energy tensor
is regular everywhere.

Let us consider a hypersurface $\Sigma $ dividing the spacetime in two regions $%
V^{\left( +\right) }$ and $V^{\left( -\right) }$. The condition
that the two metrics $g_{\mu \nu }^{\left( +\right) }$ and $g_{\mu
\nu }^{\left( -\right) }$ in the two regions $V^{\left( +\right)
}$ and $V^{\left( -\right) }$ must satisfy in order to join
smoothly on $\Sigma $ is that they must be the same on both sides
of $\Sigma $ together with their first derivatives, see for
instance Eq.(3.7.7) in \cite{poisson}, that is
\end{subequations}
\begin{equation}
g_{\mu \nu }^{\left( +\right) }|_{\Sigma }=g_{\mu \nu }^{\left( -\right)
}|_{\Sigma }~,\qquad g_{\mu \nu ,\sigma }^{\left( +\right) }|_{\Sigma
}=g_{\mu \nu ,\sigma }^{\left( -\right) }|_{\Sigma }~.  \label{junction 0}
\end{equation}%
From (\ref{junction 0}) it is therefore evident that the isochronous metric (%
\ref{NEWmetric}) satisfies such junction conditions on the hypersurfaces $%
t=t_{n}$, where it is infinitely differentiable. We also mention
that, if the junction conditions (\ref{junction 0}) on a
hypersurface $\Sigma $ are satisfied, the stress energy tensor is
regular there and it does not feature a distributional character
(thin shells) on $\Sigma $. Therefore the
stress-energy tensor associated with the isochronous metric (\ref{NEWmetric}%
) is regular everywhere.

Since the second equation (\ref{junction 0}) is expressed in terms of the
derivatives $g_{\mu \nu ,\sigma }$ of the metric tensor, which are not
tensors, such a condition is usually expressed in an invariant way by use of
the extrinsic curvature, which is diffeomorphism-invariant. Therefore (\ref%
{junction 0}) can be recast as
\begin{equation}
g_{\mu \nu }^{\left( +\right) }|_{\Sigma }=g_{\mu \nu }^{\left( -\right)
}|_{\Sigma }~,\qquad K_{ab}^{\left( +\right) }|_{\Sigma }=K_{ab}^{\left(
-\right) }|_{\Sigma }~,  \label{junction}
\end{equation}%
where $K_{ab}$ is the extrinsic curvature of $\Sigma $ defined as
\begin{equation}
K_{ab}=n_{\alpha ;\beta }e_{a}^{\alpha }e_{b}^{\beta }~,
\label{extrinsic curvature 0}
\end{equation}%
with $n^{\alpha }$ the unitary normal vector to $\Sigma $ and $e_{a}^{\alpha
}$ the three unitary tangent vectors to $\Sigma $, so that $g_{\alpha \beta
}=n_{\alpha }n_{\beta }-e_{\alpha a}e_{\beta b}\delta ^{ab}$, where $%
n_{\alpha }=g_{\alpha \beta }n^{\beta }$ and $e_{\alpha a}=g_{\alpha \beta
}e_{a}^{\beta }$. Note that (\ref{junction 0}) is more general than (\ref%
{junction}), since it is valid even where the extrinsic curvature does not
exist.

Let us now show that the second equation (\ref{junction}) is also satisfied
and the extrinsic curvature is continuous on $t=t_{n}$. The normal and
tangent vectors to the hypersurfaces $t=const$ with $t\neq t_{n}$ are $%
n^{\alpha }=\delta ^{\alpha 0}/b(t)$ and $e^{\alpha a}=-\delta ^{\alpha
a}/a(t)$, so that $n_{\alpha }=b(t)\delta _{\alpha 0}$ and $e_{\alpha
a}=a(t)\delta _{\alpha a}$. Furthermore, since $\Gamma _{\alpha \beta
}^{0}=\delta _{\alpha \beta }\,a(t)\dot{a}(t)/b(t)^{2}$ one finds
\begin{equation}
n_{\alpha ;\beta }=n_{\alpha ,\beta }-\Gamma _{\alpha \beta }^{\sigma
}n_{\sigma }=\delta _{\alpha 0}\partial _{\beta }b(t)-\Gamma _{\alpha \beta
}^{0}n_{0}=\delta _{\alpha 0}\delta _{\beta 0}\dot{b}(t)-\delta _{\alpha
\beta }\frac{\dot{a}a}{b}~.  \label{extrinsic curvature 1}
\end{equation}%
Therefore from (\ref{extrinsic curvature 0}-\ref{extrinsic curvature 1}),
using (\ref{repar}) and (\ref{definitions}) which gives $\dot{a}(t)=\alpha
^{\prime }(\tau (t))~b(t)$, one has
\begin{equation}
K_{ab}=-\delta _{ab}~\frac{\alpha ^{\prime }(\tau \left[ t\right] )}{\alpha
(\tau \left[ t\right] )}~.  \label{extrinsic curvature}
\end{equation}%
Here, of course, the appended prime denotes differentiation with respect to $%
\tau $.

This expression implies three facts. Firstly, since the extrinsic curvature
is invariant under redefinition of time, it is the same as in the equivalent
FRW reference frame (as it should be). \ Secondly, the extrinsic curvature
is only apparently singular on the hypersurfaces $t=t_{n}$. In fact, even
though the normal vector $n^{\alpha }(t)=\delta ^{\alpha 0}/b(t)$ is not
defined at $t=t_{n}$ where $b(t_{n})=0$, from (\ref{extrinsic curvature})
and ( \ref{blambda}) it is clear that both limits of $K_{ab}$ for $%
t\rightarrow t_{n}^{\left( \pm \right) }$ exist, are finite and coincide
with each other, hence the extrinsic curvature is not singular at $t=t_{n}$.
And clearly also the second junction condition in (\ref{junction}) is
satisfied at $t=t_{n}$ and both the metric tensor and the stress energy
tensor are regular there (and of course elsewhere as well). Thirdly, as long
as the inequality (\ref{condition b}) holds, the extrinsic curvature (which
is proportional to the Hubble parameter of the FRW metric (\ref{FRWmetric}))
is finite for any $t$, since, in such a case, $\alpha (\tau (t))>0$
everywhere.

Even though isochronous solutions are singularity-free and possess a regular
stress-energy tensor, they are \textit{degenerate} (non invertible) on the
hypersurfaces $t=t_{n}$. However, solutions of the Einstein equations
featuring degenerate metrics are common in general relativity. In fact the
appearance of horizons of events where the metric tensor is degenerate is
quite common: for instance the Schwarzschild solution has a metric tensor
which is degenerate (and also singular) at its horizon, but it has a well
defined physical meaning there. The isochronous solutions introduced in this
paper also feature a metric that is degenerate at certain discrete times $%
t_{n}$, but they are not singular there, indeed they are differentiable and
geodesically complete everywhere, therefore they are physically meaningful
solutions of the Einstein equations.

A further observation. We have just seen that the isochronous
cosmologies introduced in this paper may provide examples of
singularity-free Einsteinian spacetimes. Since there are theorems
by Hawking and Penrose---based on rather general
hypotheses---which guarantee the existence of singularities of the
solutions of the Einstein equations \cite{MTW,Nat}, it is
reasonable to wonder how the isochronous solutions introduced in
this paper manage to be singularity-free. The answer is that these
isochronous solutions define spacetimes which are \textit{not}
globally hyperbolic, hence which do \textit{not} fulfill the
hypotheses underpinning the validity of the theorems mentioned
above. And they are \textit{not} globally hyperbolic because they
are \textit{not} stably causal, indeed they do \textit{not}
possess a smooth function of the spacetime coordinates such that
its gradient is \textit{strictly timelike} everywhere \cite{Nat}.

Finally, it is instructive to compare the isochronous metric
(\ref{NEWmetric}) with the well known G\"{o}del metric
\cite{Godel}, providing an exact solution of the Einstein
equations in the presence of a cosmological constant and a perfect
dust fluid. Just as the isochronous metric, the G\"{o}del metric
is geodesically complete hence singularity-free, and not globally
hyperbolic. The main difference is that the G\"{o}del metric
features closed time-like null curves---this being one of its most
remarkable characteristics---and this implies that in this case
causality cannot be well defined; while no such phenomenology
haunts our isochronous solutions, since all their geodesics are
open curves (spiraling in spacetime).

The physical interpretation of the G\"{o}del solution was questioned by
Einstein \cite{Einstein}; likewise, the physical interpretation of
isochronous solutions based on the metric (\ref{NEWmetric}) can be
considered a debatable issue. But---just as Einstein did not question the
fact that the G\"{o}del solution is a \textit{bona fide} solution of his
field equations \cite{Einstein}---there is in our opinion no valid
justification to deny that the isochronous metric (\ref{NEWmetric}) provides
a \textit{bona fide} solution of Einstein's equations.

\bigskip

\section{Conclusions}

We have shown that for any given FRW cosmology described by the metric (\ref%
{FRWmetric}) it is possible to find an infinite number of \textit{isochronous%
} cosmologies described by the periodic metric (\ref{NEWmetric}) and that
one can always assign the arbitrary function $b(t)$ so that the new metric (%
\ref{NEWmetric}) is singularity free and approximates the FRW metric (\ref%
{FRWmetric}) with arbitrary accuracy for an arbitrarily long time (of
course, adequately less than the isochrony period $T$, which can be, itself,
arbitrarily assigned). One might wonder how general this finding is. To
answer this question, we note that the mechanism to find isochronous
cosmologies can be applied to \textit{any} synchronous metric, i.e. to any
metric with $g_{00}=1$ and $g_{0i}=0$. Therefore, as long as one can perform
a coordinate transformation that makes a solution of the Einstein equations
synchronous, one can find an \textit{isochronous} solution of these field
equations which approximates the original solution for an \textit{%
arbitrarily long} time with \textit{arbitrary} accuracy---indeed, which is
locally (but only locally) physically identical to the original solution up
to diffeomorphic reparameterization of the time variable. Since it is
possible to write most metrics in synchronous form by a diffeomorphic change
of coordinates, this makes our finding quite general.

Finally let us repeat (see \textit{Remark 1.1}) that, while the results that
are valid for the rather general (nonrelativistic) many-body problems
mentioned at the beginning of this paper can be extended from a classical to
a quantal context in a mathematically rigorous manner \cite{CL2007}, an
analogous quantal extension of the approach introduced in this paper is
unfeasible as long as there is no mathematically rigorous theory
encompassing general relativity and quantization.

\bigskip

\section{Acknowledgments}

This paper is the outcome of a collaboration among a relativist (FB) and a
mathematical physicist (FC) who over the last few years, together with F.
Leyvraz, investigated isochronous systems and who was interested in
extending some of these findings to a cosmological context. We wish to thank
F. Leyvraz for several useful discussions and we regret that he preferred to
opt out of co-authorship; and we also like to thank A. Degasperis for
enlightening conversations. FB is grateful to G. Calcagni, G. Gonzalez, A.
Marciano, A. Melchiorri, F. Mercati, L. Modesto, Y. Rodriguez, M. Saridakis
and W. Westra for useful discussions. We also like to thank the Centro
Internacional de Ciencias (CIC) in Cuernavaca, Mexico, for the kind
hospitality during the Gathering of Scientists on "Integrable systems and
the transition to chaos" that took place there in November-December 2012,
when this paper was largely drafted. FB thanks the UIS University of
Bucaramanga for kind hospitality during the revision of this manuscript. FB
is a Marie Curie fellow of the Istituto Nazionale di Alta Matematica
Francesco Severi.

\bigskip

\section{Appendix A: Geodesics}

\label{appendix geodesics}

The equation characterizing the geodesics of a spacetime with a given metric
$g_{\alpha \beta }$ reads
\begin{equation}
\frac{d^{2}x^{\beta }}{d\lambda ^{2}}~g_{\beta \alpha }+\Gamma _{\alpha
\beta \gamma }~\frac{dx^{\beta }}{d\lambda }\frac{dx^{\gamma }}{d\lambda }%
=g_{\beta \alpha }~\frac{dx^{\beta }}{d\lambda }\frac{1}{L}\frac{d{L}}{%
d\lambda }~,  \label{geod1}
\end{equation}%
where Greek indices run from $0$ to $3$, $\Gamma _{\alpha \beta \gamma
}=(g_{\beta \alpha ,\gamma }+g_{\gamma \alpha ,\beta }-g_{\beta \gamma
,\alpha })/2$ are the Christoffel symbols, $\lambda $ is a parameter used to
parameterize the geodesic and $L=\sqrt{(dx^{\alpha }/d\lambda )(dx^{\beta
}/d\lambda )g_{\alpha \beta }}$ for timelike geodesics, $L=\sqrt{%
-(dx^{\alpha }/d\lambda )(dx^{\beta }/d\lambda )g_{\alpha \beta }}$ for
spacelike geodesics (an analogous, if not quite identical, treatment applies
to null geodesics; we omit it to avoid repetition). Note that this equation,
(\ref{geod1}), is adequate to define the geodesics even where the metric $%
g_{\alpha \beta }$ is not invertible; it is only the reduction to its more
standard version,
\begin{equation}
\frac{d^{2}x^{\alpha }}{d\lambda ^{2}}+\Gamma _{\,\,\,\beta \gamma }^{\alpha
}\frac{dx^{\beta }}{d\lambda }\frac{dx^{\gamma }}{d\lambda }=\frac{%
dx^{\alpha }}{d\lambda }\frac{1}{L}\frac{d{L}}{d\lambda }~,  \label{geod2}
\end{equation}%
that is not defined when the metric $g_{\alpha \beta }$ is not invertible.

Let us suppose that the geodesics associated with the FRW metric (\ref%
{FRWmetric}) are identified by the formula
\begin{equation}
Y=\left[ Y^{0}(\lambda ),~\vec{Y}(\lambda )\right]  \label{geodesic Y def}
\end{equation}%
where $Y^{0}$ coincides with the FRW cosmological time along the geodesics.
We use this notation to emphasize the fact that geodesics are geometric
curves and they do not depend of their parameterization. Let us specify the
geodesic equation (\ref{geod1}) (or equivalently (\ref{geod2})) in the case
of the FRW metric (\ref{FRWmetric}), which gives
\begin{subequations}
\label{geodesic Y}
\begin{equation}
\frac{d^{2}Y^{0}}{d\lambda ^{2}}+\alpha ^{\prime }(Y^{0})\alpha (Y^{0})\frac{%
dY^{i}}{d\lambda }\frac{dY^{j}}{d\lambda }\delta _{ij}=\frac{dY^{0}}{%
d\lambda }\frac{1}{L_{Y}}\frac{d{L_{Y}}}{d\lambda }~,  \label{geodesic Y 1}
\end{equation}%
\begin{equation}
\frac{d^{2}Y^{i}}{d\lambda ^{2}}+\frac{\alpha ^{\prime }(Y^{0})}{\alpha
(Y^{0})}\frac{dY^{i}}{d\lambda }\frac{dY^{0}}{d\lambda }=\frac{dY^{i}}{%
d\lambda }\frac{1}{L_{Y}}\frac{d{L_{Y}}}{d\lambda }~,  \label{geodesic Y 2}
\end{equation}%
where Latin indices run from $1$ to $3$, $\alpha ^{\prime }(Y^{0})\equiv
d\alpha (Y^{0})/dY^{0}$ and \newline
$L_{Y}=\sqrt{|(dY^{\alpha }/d\lambda )(dY^{\beta }/d\lambda )g_{\alpha \beta
}|}$ with $g_{\alpha \beta }$ given by (\ref{FRWmetric}) for timelike and
spacelike geodesics.

Likewise the geodesics associated with the isochronous metric (\ref%
{NEWmetric}) are given by
\end{subequations}
\begin{equation}
X=\left[ X^{0}(\lambda ),~\vec{X}(\lambda )\right]  \label{geodesic X def}
\end{equation}%
where again $X^{0}$ coincides with the time along the geodesics of the
metric (\ref{NEWmetric}). The geodesic equations for the metric (\ref%
{NEWmetric}) are
\begin{subequations}
\label{geodesic X}
\begin{equation}
\frac{d^{2}X^{0}}{d\lambda ^{2}}+\frac{b^{\prime }(X^{0})}{b(X^{0})}\left(
\frac{dX^{0}}{d\lambda }\right) ^{2}+\frac{a^{\prime }(X^{0})a(X^{0})}{%
b^{2}(X^{0})}\frac{dX^{i}}{d\lambda }\frac{dX^{j}}{d\lambda }\delta _{ij}=%
\frac{dX^{0}}{d\lambda }\frac{1}{L_{X}}\frac{d{L_{X}}}{d\lambda }~,
\label{geodesic X 1}
\end{equation}%
\begin{equation}
\frac{d^{2}X^{i}}{d\lambda ^{2}}+\frac{a^{\prime }(X^{0})}{a(X^{0})}\frac{%
dX^{i}}{d\lambda }\frac{dX^{0}}{d\lambda }=\frac{dX^{i}}{d\lambda }\frac{1}{%
L_{X}}\frac{d{L_{X}}}{d\lambda }~,  \label{geodesic X 2}
\end{equation}%
where $a^{\prime }(X^{0})\equiv da(X^{0})/dX^{0}$ and $L_{X}=\sqrt{%
|(dX^{\alpha }/d\lambda )(dX^{\beta }/d\lambda )g_{\alpha \beta }|}$ with $%
g_{\alpha \beta }$  the isochronous metric (\ref{NEWmetric}) for
timelike and spacelike geodesics.

Using the fact that $a(X^{0})\equiv \alpha (B(Y^{0}))$ (see (\ref%
{definitions})), it is easy to see that the geodesics (\ref{geodesic Y def})
and (\ref{geodesic X def}), solutions of the systems (\ref{geodesic Y}) and (%
\ref{geodesic X}) respectively, are related by the following equalities
\end{subequations}
\begin{equation}
\frac{dY^{0}(\lambda )}{d\lambda }=b(X^{0}(\lambda ))\,\,\frac{%
dX^{0}(\lambda )}{d\lambda };\quad \frac{dY^{i}(\lambda )}{d\lambda }=\frac{%
dX^{i}(\lambda )}{d\lambda }~.  \label{geodesic X-Y}
\end{equation}%
These relations can be integrated to give (up to $4$ irrelevant integration
constants)%
\begin{equation}
Y^{0}(\lambda )=B(X^{0}(\lambda ));\quad Y^{i}(\lambda )=X^{i}(\lambda )~.
\label{geodesic X-Y}
\end{equation}%
If the geodesics of the metric (\ref{FRWmetric}) are parameterized by $%
Y^{0}=\lambda =\tau $ and
\begin{equation}
Y=\left[ \lambda ,~\vec{Y}\left( \lambda \right) \right] ~,
\end{equation}%
from (\ref{geodesic X-Y}), using the fact that $X^{i}(\lambda
)=Y^{i}(Y^{0}=\lambda )=Y^{i}(Y^{0}=B(X^{0}(\lambda )))$ and
re-parameterizing the geodesic $X$ by use of its proper time $X^{0}\equiv
\mu =t$ instead of the parameter $\lambda $, one has that the geodesics of
the isochronous metric (\ref{NEWmetric}) are given by%
\begin{equation}
X=\left[ \mu ,~\vec{Y}\left( B(\mu )\right) \right] =\left[ t,~\vec{Y}\left(
\tau (t)\right) \right] ~,
\end{equation}%
which is our final result, see (\ref{Geodesic isochronous 0}) and (\ref%
{condition b}).

For instance, one can easily verify that the curves

\begin{equation}
X=\left[ X^{0}=\lambda ,~\vec{X}(\lambda )=\vec{X}(0)\right]
\label{geodesic X solutions}
\end{equation}%
are time-like geodesics of the metric (\ref{NEWmetric}), solutions of (\ref%
{geodesic X 1}-\ref{geodesic X 2}), which correspond to observers at rest in
the reference frame of (\ref{NEWmetric}). They pass through the
hypersurfaces $t=t_{n}$ smoothly, therefore no singularity can be present
there.

\end{document}